\documentclass[superscriptaddress,nofootinbib,twocolumn,floatfix,notitlepage]{revtex4-1}
\usepackage{graphicx,epstopdf}
\usepackage{amsmath,amssymb,graphicx}
\usepackage[usenames, dvipsnames]{color}
\usepackage{times}
\usepackage{array}
\usepackage{latexsym}
\usepackage{floatrow}
\usepackage{longtable}
\usepackage[utf8, latin9]{inputenc}
\usepackage{tikz,hyperref}
\definecolor{steelblue}{RGB}{25,25,112}
\hypersetup{colorlinks,linkcolor={blue},citecolor={blue},urlcolor={steelblue}}

\definecolor{darkyellow}{rgb}{0.545098,0,0}
\definecolor{teal}{rgb}{0,0.5,0.5}

\newcommand{\blue}{}

\newcommand{\Msun}{M_\odot}
\newcommand{\kmps}{{\rm km\,s^{-1}}}

\newcommand{\Vmax}{V_{\rm max}}

\newcommand{\rcut}{r_{\rm\textup{cut}}}

\newcommand{\Msub}{M_{\rm \textup{sub}}}
\newcommand{\rsub}{r_{\rm \textup{sub}}}

\newcommand{\MBH}{M_{\rm BH}}
\newcommand{\Mh}{M_{\rm halo}}

\newcommand{\rsp}{r_{\rm \textup{sp}}}
\newcommand{\rin}{r_{\rm \textup{in}}}

\newcommand{\orcid}[1]{\href{https://orcid.org/#1}{\includegraphics[width=8pt]{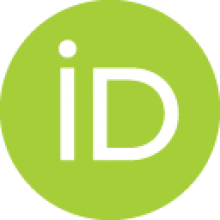}}}

\begin{document}

\title{SIDM and CDM interpretations of the million-solar-mass lensing perturber JVAS B1938+666-$\mathcal{V}$}

\author{Xingyu Zhang\orcid{0009-0009-2791-1684}}
\email{xingyuz@ucr.edu}
\affiliation{Center for Experimental Cosmology \& Instrumentation, Department of Physics and Astronomy, University of California, Riverside, California 92521, USA}

\author{Hai-Bo Yu\orcid{0000-0002-8421-8597}}
\email{haiboyu@ucr.edu}
\affiliation{Center for Experimental Cosmology \& Instrumentation, Department of Physics and Astronomy, University of California, Riverside, California 92521, USA}

\begin{abstract}

{
A $10^6\,\Msun$ object has recently been inferred from gravitational imaging of the strong-lensing system JVAS B1938+666, exhibiting an unusually dense inner region embedded within an extended envelope, far exceeding expectations for cold dark matter (CDM) halos. Using gravothermal fluid simulations, we show that such a structure arises naturally in self-interacting dark matter (SIDM) halos evolving into a deep core-collapse phase, where an ultra-dense core forms within an extended profile. The resulting density structure closely matches the inferred properties of the lensing object. We also demonstrate that a similar profile could be reproduced in CDM in the presence of an intermediate-mass black hole, but this requires an early-forming progenitor that subsequently loses $5$ orders of magnitude in mass through tidal stripping by the lens galaxy. Whether such a scenario can be realized in realistic cosmological environments remains an open question.
}

\end{abstract}
\maketitle

\section{Introduction}
\label{sec:intro}

Strong gravitational lensing is emerging as a powerful probe of otherwise invisible structure on sub-galactic scales~\cite{Vegetti:2023mgp,Gannon:2025nhr}. A recent highlight is the detection of a $\sim10^6\,\Msun$ perturber in the lens system JVAS B1938+666 at redshift $z=0.881$ \cite{Powell:2025rmj}. The inferred mass is highly concentrated, with $40\%$ enclosed within a projected radius of $80\,\mathrm{pc}$. Such a high mass concentration is highly atypical for a (sub)halo in the cold dark matter (CDM) framework, corresponding to a $9\sigma$ outlier in the concentration--mass distribution \cite{Yu:2025tmp}. By contrast, this feature can arise naturally in self-interacting dark matter (SIDM)~\cite{Tulin:2017ara,Adhikari:2022sbh}, where halos can undergo gravothermal collapse, leading to a dramatic increase in central density. Ineed, Ref.~\cite{Yu:2025tmp} showed that the inferred density profile of the perturber is well reproduced by core-collapsed SIDM halos simulated in~\cite{Zhang:2024fib}, with implications for other dense perturbers inferred from stellar streams \cite{Bonaca:2018fek} and satellite galaxies~\cite{Penarrubia:2024vms}.

More recently, Ref.~\cite{Vegetti:2026mmx} further explored a broader set of mass models and redshift configurations for the perturber beyond the previous analysis~\cite{Powell:2025rmj}. A favored interpretation is that it represents a substructure embedded within the main lens galaxy, comprising an unresolved, point-like component confined within $10\,{\rm pc}$ with a mass of a few times $10^5\,\Msun$, surrounded by a more extended distribution. In this scenario, more than $10\%$ of the total mass is concentrated within the inner $10\,{\rm pc}$, corresponding to {\blue a} $20\sigma$ outlier in the CDM framework~\cite{Vegetti:2026mmx}. Importantly, a purely point-like mass distribution is also strongly disfavored.

In this work, we show that the refined mass distribution of the lensing perturber reported in~\cite{Vegetti:2026mmx} is a characteristic prediction of the SIDM framework. As an SIDM halo enters the gravothermal collapse phase, it develops an extremely dense central core embedded within a smooth extended profile; this inner core can exhibit a point-like component. Using fluid simulations, we resolve the formation of this core and follow its evolution in detail, overcoming the resolution limitations of the $N$-body approach adopted in~\cite{Yu:2025tmp}. For initially CDM-like progenitor halos with masses of $\sim10^{8}\,\Msun$, moderately high concentrations, and self-interaction cross sections of $\sim100\,\mathrm{cm^2\,g^{-1}}$, we find that gravothermal evolution, combined with mild tidal mass loss, can drive the cylindrical mass within $10\,\mathrm{pc}$ to $3\times10^5\,\Msun$, producing a mass profile that closely matches the inferred model of the perturber in~\cite{Vegetti:2026mmx}.

We also propose an alternative CDM-based interpretation, in which the perturber is a tidally stripped remnant of a $10^{11}\,\Msun$ progenitor hosting a $1.9\times10^5\,\Msun$ central black hole. The black hole deepens the gravitational potential and induces the formation of a central density spike, reproducing the inner mass distribution of the lensing object reported in~\cite{Vegetti:2026mmx}. In this CDM scenario, an extreme mass loss of $\sim5$ orders of magnitude is required, making it potentially difficult to realize in realistic cosmological environments.

The rest of this paper is organized as follows. In Sec.~\ref{sec:sidm}, we present fluid simulations for the SIDM interpretation and compare the resulting mass profiles with the lensing measurements. In Sec.~\ref{sec:cdm}, we develop a CDM model in which the perturber is a tidally stripped remnant hosting an intermediate-mass black hole and assess its consistency with the inferred mass distribution. We conclude in Sec.~\ref{sec:con}. In Appendix~\ref{sec:tidal}, we examine the orbital requirements needed to produce the extreme tidal stripping invoked in the CDM scenario.

\section{The SIDM scenario}

\label{sec:sidm}

\begin{figure*}[t]
	\centering
	\includegraphics[scale=0.65]{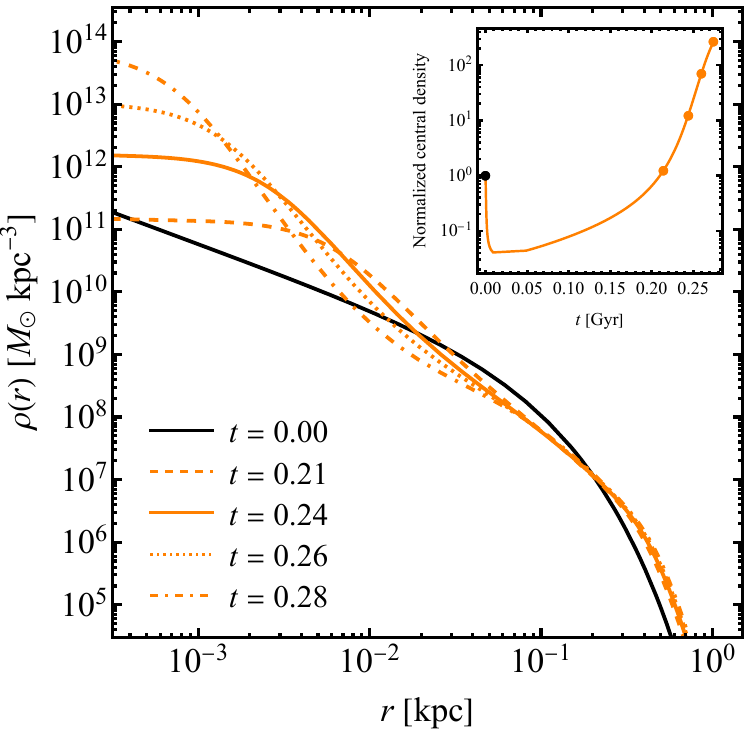}~~~~~~~~
	\includegraphics[scale=0.65]{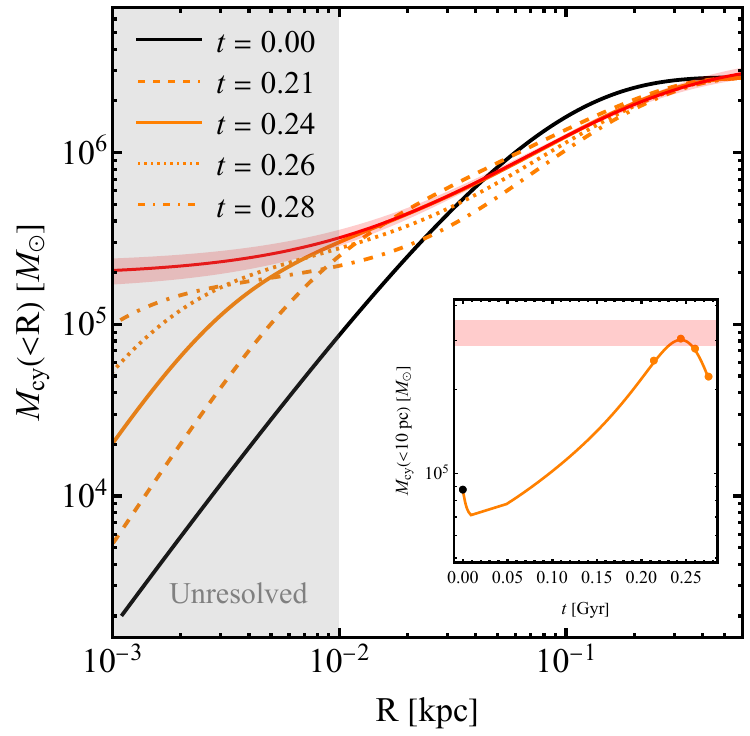}
	\caption{{\it Left panel}: Density profiles at selected snapshots during the collapse phase (orange) for $t=0.21\,{\rm Gyr}$ (dashed), $0.24\,{\rm Gyr}$ (solid), $0.26\,{\rm Gyr}$ (dotted), and $0.28\,{\rm Gyr}$ (dot-dashed), together with the truncated NFW profile (solid black). The concentration of the initial halo is $c_{200}=50$. The inset shows the full evolution of the normalized central density, with circles marking the SIDM snapshots displayed in the main panel. {\it Right panel}: The corresponding projected cylindrical mass profiles of the simulated SIDM halo at the five snapshots, together with the ``Pseudo-Jaffe+Point Mass'' (PJ+PM) model for the perturber (red)~\cite{Vegetti:2026mmx}, where the shaded band denotes the $1\sigma$ uncertainty. The best-matching snapshot is $t=0.24\,{\rm Gyr}$. The unresolved region $R<10\,\mathrm{pc}$ is shaded in gray. The inset shows the evolution of the cylindrical mass within $10 \, {\rm pc}$. Horizontal red band denotes the $1\sigma$ interval of the PJ+PM model. Filled circles mark the snapshots shown in the main panel.}
	\label{fig-SIDM}
\end{figure*}

For the SIDM scenario, we consider a progenitor halo with an initial Navarro--Frenk--White (NFW) density profile~\cite{Navarro:1996gj} 
\begin{equation}
	\rho_{\rm NFW}(r) = \frac{\rho_s}{\left( r/r_s \right)(1+r/r_s)^2},
	\label{eq:nfw}
\end{equation}
where $r_s$ and $\rho_s$ are scale radius and density, respectively. In our fiducial model, the halo has a mass of $M_{200}=1.5\times10^8\,\Msun$ and a concentration of $c_{200}=50$, corresponding to a $2.8\sigma$ upward fluctuation relative to the median concentration--mass relation~\cite{Diemer:2018vmz}.  Accordingly, $r_s=0.22\,\mathrm{kpc}$, $\rho_s=3.7\times10^8\,\Msun\,\mathrm{kpc^{-3}}$, and the maximum circular velocity is $V_{\rm max}=15\,{\rm km\,s^{-1}}$. We have verified that these initial conditions are consistent with the high-concentration tail of halos in the Concerto suite of cosmological zoom-in simulations~\cite{Nadler:2025jwh} (see Fig.~2 of~\cite{Kong:2025sqx}).

Since the perturber is a substructure of the main lens galaxy, we add a truncation to the initial NFW profile following the analytical form proposed in~\cite{Errani:2020wgn},
\begin{equation}
	\rho_{\rm NFW}(r) \times \frac{e^{-r/\rcut}}{(1+{r_s}/{\rcut})^{0.3}},
	\label{eq:tnfw}
\end{equation}
where the truncation radius is set to be $\rcut = 0.1 \, {\rm kpc}$. With the truncation, the halo mass is reduced to $2.7\times10^6 \, \Msun$, comparable to that inferred for the perturber, and the maximum circular velocity is $\Vmax=7.3 \, \mathrm{km\,s^{-1}}$. For the self-interaction cross section, we adopt a benchmark value of $\sigma/m=100 \, \mathrm{cm^2 \, g^{-1}}$, which is broadly consistent with the effective cross section on the relevant mass scale in the velocity-dependent SIDM models simulated in the Concerto suite~\cite{Nadler:2025jwh}. To model the gravothermal evolution, we have developed an efficient \texttt{C} code based on the public \texttt{Python} code developed in~\cite{Outmezguine:2022bhq,Gad-Nasr:2023gvf}, which is built upon the gravothermal fluid formalism~\cite{Pollack:2014rja,Balberg:2002ue}. Our code also incorporates initial conditions based on the truncated NFW profile as in Eq.~\ref{eq:tnfw}. 

In Fig.~\ref{fig-SIDM} (left panel), we show the density profiles of the simulated halo at four snapshots during the core-collapse phase $t=0.21\textup{--}0.28\,{\rm Gyr}$ (orange) together with the truncated NFW profile (black). The inset shows the full evolution of the normalized central density, with the circles marking the snapshots displayed in the main panel. It is clear that an ultra-dense core emerges for $t\gtrsim0.21\,{\rm Gyr}$ and its central density grows fast. Among the four collapsed profiles, the evolution time differs by only $0.07\,{\rm Gyr}$ between the lowest-density (dashed) and highest-density (dot-dashed) snapshots, while the central density increases by a factor of $300$, demonstrating that gravothermal collapse is indeed a runaway process. Within the inner region $r<10 \, \mathrm{pc}$, the collapsed halo develops substantially higher densities than the truncated NFW profile. In the outer region $r\gtrsim0.2\,\mathrm{kpc}$, the SIDM halo exhibits slightly enhanced densities due to the kick-out effect~\cite{Kong:2024zyw}. 

For the initial progenitor halo prior to tidal truncation, we estimate the core-collapse timescale as~\cite{Essig:2018pzq}
\begin{equation}
	\label{eq-tc}
		t_c = \frac{200}{r_s \rho_s\left(\sigma_{\mathrm{eff}} / m\right)} \frac{1}{\sqrt{4 \pi G \rho_s}} = 0.82\,{\rm Gyr},
\end{equation}
which is approximately a factor of three longer than the collapse time obtained from the fluid simulations with the truncated NFW profile. This difference arises because tidal truncation accelerates the onset of gravothermal collapse by reducing the peak of the dark matter velocity-dispersion profile and enhancing the dispersion gradient~\cite{Nishikawa:2019lsc,Sameie:2019zfo,Kahlhoefer:2019oyt,Yang:2021kdf,Zeng:2021ldo}.

{\blue Figure~\ref{fig-SIDM} (right, main panel) shows the corresponding projected cylindrical mass profiles at the five snapshots. The mass profile at $t = 0.24\,{\rm Gyr}$ best matches the  ``Pseudo Jaffe+Point Mass'' (PJ+PM) model of the perturber (red)~\cite{Vegetti:2026mmx}.} For the region $\gtrsim10 \, \mathrm{pc}$, owing to the formation of the ultra-dense core, the mass is sufficiently large to account for the unresolved point-mass-like component inferred in~\cite{Vegetti:2026mmx}. Within the unresolved region {\blue$R<10\,\mathrm{pc}$} (shaded gray), the best-matching SIDM mass profile deviates from the PJ+PM model. This difference is expected because the model assumes a point-like central mass, yielding a flattened mass profile toward the center, whereas the central core in the collapsed SIDM halo is not point-like and the cylindrical mass decreases toward smaller radii. However, since this inner region is not resolved by current observations~\cite{Vegetti:2026mmx}, the two scenarios cannot yet be distinguished observationally.

\begin{figure}[h]
	\centering
	\includegraphics[scale=0.65]{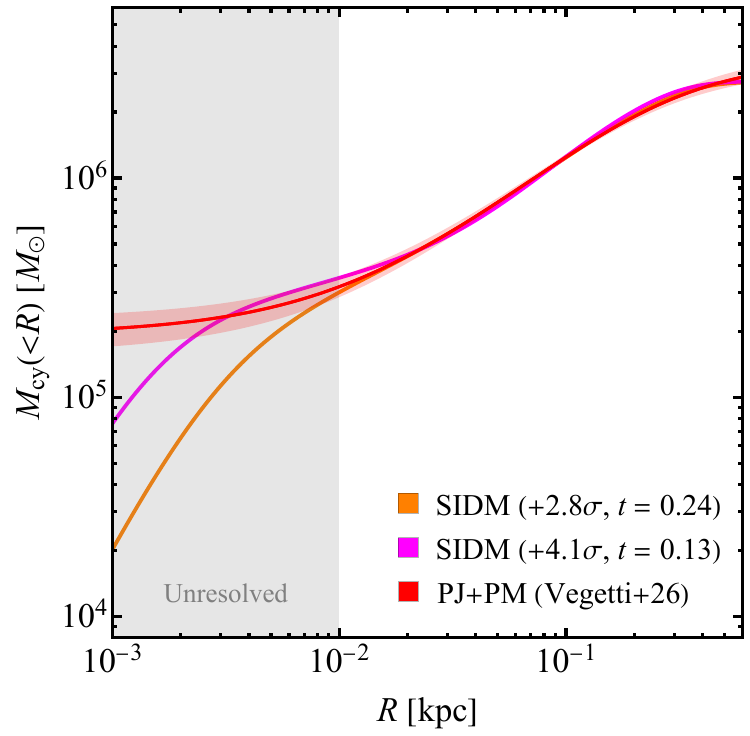}
	\caption{The projected cylindrical mass profile for another simulated SIDM halo at the snapshot $t=0.13\,{\rm Gyr}$ (magenta); its initial concentration is $c_{200} = 86$, corresponding to a $4.1\sigma$ upward fluctuation relative to the median. For comparison, the fiducial model (orange) and the PJ+PM model (red)~\cite{Vegetti:2026mmx} are also shown. The unresolved region $R<10\,\mathrm{pc}$ is shaded in gray.}
	\label{fig-Mcy}
\end{figure}

{ 
\blue 

As the halo evolves beyond $t=0.24\,{\rm Gyr}$, the cylindrical mass within a few ${\rm pc}$ increases, while that at larger radii decreases with time. After the best-matching snapshot (solid), the mass profile over $R=10\textup{--}30\,{\rm pc}$ falls below the inferred mass model. This reflects the gravothermal evolution of the SIDM halo: although the central core density continues to increase, high-velocity dark matter particles near the surface of the core tend to migrate to the outer halo, producing a sharp density drop. Thus, there is an optimal time window during which the SIDM halo remains compatible with the observations. Figure~\ref{fig-SIDM} (right, inset) shows the cylindrical mass enclosed within $10\,{\rm pc}$ as a function of evolution time. $M_{\rm cy}(<10\,{\rm pc})$ reaches a maximum at $t\approx0.24\,{\rm Gyr}$, corresponding to the best-matching snapshot, and subsequently decreases. During the time window $t=0.23\textup{--}0.26\,{\rm Gyr}$, the simulated halo remains consistent with the PJ+PM model (red band). Therefore, the snapshot that best matches the current observations corresponds to a transient stage of gravothermal evolution. For a given subhalo, some fine-tuning is required for it to be observed within the optimal time window. However, the lens galaxy is expected to host a large population of low-mass subhalos, making it plausible that some of them are at the appropriate evolution stage.}

{\blue To further investigate whether a higher halo concentration can better reproduce the point-mass-like component, we perform an additional simulation with {\blue $M_{200}= 5.6 \times10^7 \, \Msun$, $c_{200}=86$ ($+4.1\sigma$)}, and $\rcut=0.1\,\mathrm{kpc}$. Figure~\ref{fig-Mcy} shows the projected cylindrical mass profile of the simulated halo at the best-matching snapshot $t=0.13\,\mathrm{Gyr}$ (magenta). We find that the model reproduces the PJ+PM profile well down to $R\approx3\,\mathrm{pc}$. However, even for the high concentration, the cylindrical mass profile decreases toward the center, in contrast to the flattened profile expected for a model containing a point mass. We also note that the two SIDM halo models are nearly indistinguishable over the region currently resolved by observations, indicating a degeneracy between the halo parameters and evolutionary stage. Nevertheless, they differ significantly within the unresolved region $R<10\,{\rm pc}$. Future observations with improved spatial resolution could break this degeneracy and provide a more stringent test of the SIDM interpretation.}

We note that the perturber model without a point-mass component from the earlier analysis~\cite{Powell:2025rmj} has a lower cylindrical mass within $10\,\mathrm{pc}$ by a factor of $1.8$ compared to the PJ+PM model~\cite{Vegetti:2026mmx}. Ref.~\cite{Yu:2025tmp} showed that this earlier model can be well reproduced by a core-collapsed SIDM halo simulated using the $N$-body approach in~\cite{Zhang:2024fib}, which lacks the resolution required to fully capture the formation of the ultra-dense core. In contrast, the fluid approach adopted here achieves much higher spatial resolution, allowing us to resolve the formation and evolution of the core in detail. Although the choice of halo parameters is not unique because of degeneracies among the halo mass, concentration, truncation radius, and self-interaction cross section, we find that simultaneously reproducing both the large cylindrical mass $3\times10^5\,\Msun$ within $R=10\,\mathrm{pc}$ and the overall perturber mass of $\sim10^6\,\Msun$ significantly constrains the allowed parameter space and motivates the halo parameters adopted in this work.

An interesting question is whether the point-like component of the perturber could originate from a black hole formed through SIDM core collapse. In the absence of a strong baryonic potential, recent studies have shown that the region subject to general-relativistic instability~\cite{Feng:2020kxv,Feng:2021rst} has a characteristic mass of $M_{\rm seed}\sim4\times10^{-6}\,M_{\rm SMFP}$~\citep{Gu:2026zzq,Feng:2025rzf}, where $M_{\rm SMFP}$ denotes the mass enclosed within the short-mean-free-path (SMFP) region. For the fiducial parameters adopted here, at the snapshot $t=0.24\,\mathrm{Gyr}$, the SMFP radius is $15\,\mathrm{pc}$ and the enclosed mass is $M_{\rm SMFP}=3.1\times10^5\,\Msun$. This corresponds to a seed black hole mass of only $M_{\rm seed}\approx1.2\,\Msun$. If the entire SMFP region eventually collapses into a black hole through dark accretion~\cite{Feng:2025rzf}, it could produce a black hole with a mass comparable to the point-like mass component as in~\cite{Vegetti:2026mmx}. Future observations capable of resolving the inner region $R<10\,\mathrm{pc}$ could provide a more stringent test of this scenario and potentially distinguish between a dense SIDM core and a central black hole.

\section{The CDM scenario}

\label{sec:cdm}

\begin{figure*}[t]
	\centering
	\includegraphics[scale=0.38]{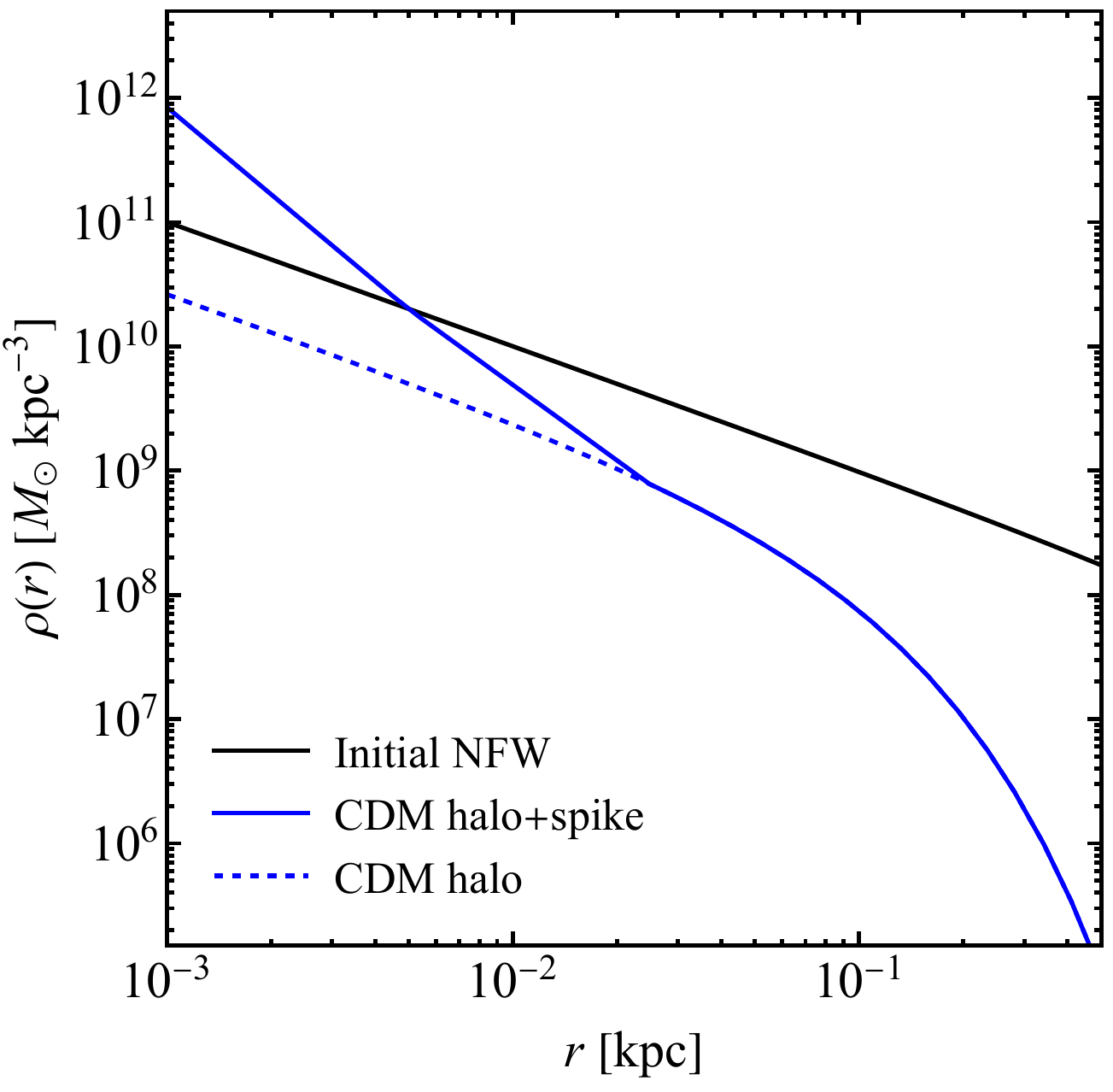}~~
	\includegraphics[scale=0.38]{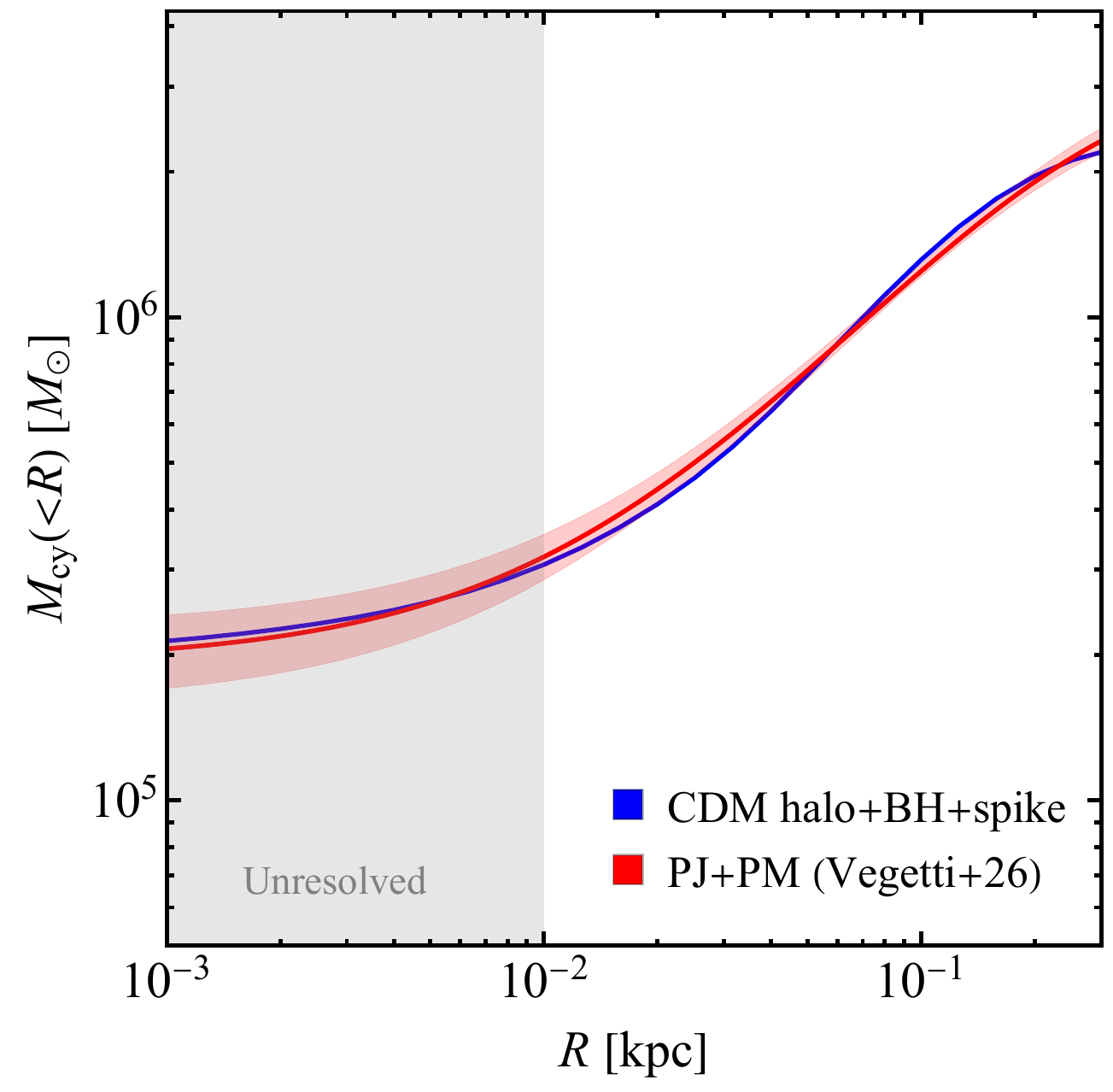}
	\caption{
	{\it Left panel}: Density profiles for the truncated CDM halo hosting a central black hole, shown with (solid blue) and without (dashed blue) the black hole-induced density spike, together with the initial NFW halo (solid black). {\it Right panel}: Projected cylindrical mass profile for the tidal remnant (truncated CDM halo+black hole+spike; solid blue), together with the PJ+PM model (red)~\cite{Vegetti:2026mmx}, where the shaded band denotes the $1\sigma$ uncertainty. The unresolved region $R<10\,{\rm pc}$ is shaded in gray. }
	\label{fig-CDM}
\end{figure*}

As discussed previously, a CDM halo is not sufficiently compact to explain the high central density of the perturber, particularly its point-like mass component. Nevertheless, the perturber could potentially originate from the remnant of a massive progenitor hosting a $\sim10^5\,\Msun$ central black hole that subsequently loses most of its mass through tidal stripping by the lens galaxy. In the presence of a central black hole, a dark matter density spike may form, producing a profile substantially steeper than the NFW cusp~\cite{Gondolo:1999ef} and thereby alleviating the tension with the inferred inner mass distribution. In this section, we develop a mass model for this scenario.  

We follow the relation between black hole mass $\MBH$ and host halo mass $\Mh$ as~\citep{Bandara:2009sd, Booth:2009zb},
\begin{equation}
	\label{eq-MBH-Mhalo}
	\begin{split}
		\log_{10}\left( \frac{\MBH}{\Msun} \right)  = \alpha + \beta \left[\log_{10}\left(\frac{\Mh}{\Msun}\right) - 13\right],
	\end{split}
\end{equation}
where $\alpha = (8.18 \pm 0.11)$ and $\beta = (1.55 \pm 0.31)$. From Eq.~\ref{eq-MBH-Mhalo}, we find that the expected halo mass is $10^{11}\,\Msun$ for a $10^{5}\,\Msun$ black hole. More {\blue specifically}, we adopt $\Mh=M_{200} = 10^{11} \, \Msun$ and $\MBH = 1.9 \times 10^5 \, \Msun$ in our numerical calculation. The halo concentration is $c_{200} = 14.6$, corresponding to a $0.9\,\sigma$ upward fluctuation relative to the median~\cite{Diemer:2018vmz}. Assuming an NFW profile for the halo (see Eq.\ref{eq:nfw}), the scale density and radius are $\rho_s = 1.5 \times 10^7\,\Msun\,{\rm kpc^{-3}}$ and $r_s = 6.6\,{\rm kpc}$, respectively; $V_{\rm max}= 88 \, \kmps$.

For the CDM density spike, we assume that the black hole grows adiabatically and the logarithmic density slope of the spike is given by $\gamma_{\rm sp}={(9-2\gamma)}/{(4-\gamma)}$, where $\gamma=1$ for an NFW profile and hence {\blue $\gamma_{\rm sp} = 7/3$}. We determine the spike radius $\rsp = 0.2 \, \rin$, where $\rin$ is the radius of influence of the black hole, calculated through the condition $ \int_{0}^{\rin} 4\pi r^2 \rho_{\rm NFW}(r) dr= 2\MBH$~\cite{Merritt:2003qc}, where $\rho_{\rm NFW}$ is the initial halo density without a spike. For the model parameters we take, we obtain $\rin = 25\,{\rm pc}$ and $\rsp = 5 \, {\rm pc}$. As the progenitor falls into the potential well of the lens galaxy, its outerskirt will be heavily stripped. Nevertheless, we assume that the black hole dominates the dynamics so that the density within the spike radius is intact. This is a reasonable assumption given the fact that $\rsp$ is less than $10\,{\rm pc}$. 

For the region $r > \rin$, where the black hole's influence is diminishing, we apply a truncated NFW profile to the halo by assuming that it follows the tidal track~\cite{Errani:2020wgn, Errani:2021rzi}. For the intermediate region $\rsp<r<\rin$, we interpolate the spike and truncated density profiles. The resulting overall density profile of the tidal remnant of a spike\textup{--}halo system is therefore
\begin{equation}
	\label{eq-rhoall}
	\rho(r) = 
	\begin{cases}
		\rho_{\rm NFW}(\rsp)\left( \frac{r}{\rsp} \right)^{-7/3}, & 0< r \leqslant \rsp \\[10pt]
		\rho_{\rm NFW}(\rsp) \left( \frac{r}{\rsp} \right)^\lambda, & \rsp < r < \rin,\\[10pt]
		\rho_{\rm NFW}(r) \times \frac{e^{-r/\rcut}}{\left(1+{r_s}/{\rcut}\right)^{0.3}}, & r\geqslant \rin 
	\end{cases}
\end{equation}
where $\rho_{\rm NFW}(r)$ is the initial density profile before tidal stripping and $r_s$ is its scale radius, $\lambda= [\log\rho(\rin)-\log\rho(\rsp)]/[{\log \rin - \log \rsp}]$, and $\rcut$ the truncation radius, which we take to be $80 \,{\rm pc}$. This value is consistent with the expectation from the tidal-track relation, where the truncation radius is set by the ratio of initial to final masses of the progenitor; see Eq. 9 of~\cite{Errani:2020wgn}.

In Fig.~\ref{fig-CDM} (left panel), we show the density profiles for the truncated NFW halo hosting a central black hole, with (solid blue) and without (dashed blue) the black hole-induced density spike. The spike substantially enhances the dark matter density within $10\,\mathrm{pc}$. Relative to the initial NFW halo (solid black), the density of the truncated halo is exponentially suppressed in the region $r\gtrsim0.1\,\mathrm{kpc}$. The total mass of the remnant is $2.1 \times 10^6\,\Msun$ (black hole+truncated NFW halo+spike), consistent with that inferred for the perturber. Fig.~\ref{fig-CDM} (right panel) shows that the cylindrical mass profile of the remnant (solid blue) agrees well with the PJ+PM model (red) from~\cite{Vegetti:2026mmx} over the entire radial range shown. The density spike contributes $30\%$ of the cylindrical mass within $R=10\,\mathrm{pc}$. Without the spike, the combination of the black hole and truncated NFW halo remains too shallow to reproduce the high central density of the perturber, requiring a higher halo concentration.

A potential challenge for the CDM scenario is whether the progenitor halo can undergo the substantial mass loss required by the observations. The initial halo mass must be $\sim 10^{11}\,\Msun$ to host a $10^5\,\Msun$ black hole, whereas the final halo mass must be reduced to $\sim 10^6\,\Msun$ to match the inferred observational value. To investigate the feasibility of such extreme tidal stripping, we employ a semi-analytic method~\cite{Du:2024sbt} assuming a circular orbit. We find that sufficient mass loss within $6\,{\rm Gyr}$ can be achieved only if the orbital radius is smaller than $\sim 30\,{\rm kpc}$ from the center of the host halo; see Appendix~\ref{sec:tidal} for details. Such small orbital radii imply that the progenitor must have fallen into the host at an early cosmic time. Therefore, the CDM scenario requires two conditions to be satisfied simultaneously: first, the formation of a $10^{11}\,\Msun$ halo hosting a $10^5\,\Msun$ central black hole at very early times; and second, the rapid infall of this halo shortly after its formation. Further work is needed to determine whether these conditions can be realized in the early Universe. 

{\blue
Another interesting possibility is that the perturber is a $10^5\,\Msun$ primordial black hole (PBH). In this case, the black hole--halo mass relation does not directly apply, potentially alleviating the challenges discussed above. The abundance of $10^5\,\Msun$ PBHs is subject to stringent constraints. Assuming Gaussian primordial fluctuations, the PBH abundance relative to dark matter is constrained to be $<10^{-10}$ by CMB spectral-distortion observations~\cite{Carr:2020gox}. This bound can be relaxed if the primordial fluctuations are non-Gaussian and the PBHs form through mergers~\cite{Taniguchi:2000mp,Mouri:2002mc,Clesse:2015wea,Zhang:2025tgm,Zhang:2026vjk}. A more robust bound is $\lesssim10^{-4}$ from X-ray binary accretion constraints~\cite{Carr:2020gox}. A PBH is also expected to seed the growth of a surrounding dark matter halo~\cite{Ricotti:2009bs}. An isolated $10^5\,\Msun$ PBH can develop a halo with a mass of $2\times10^8\,\Msun$ and a characteristic radius of $6\,{\rm kpc}$ by $z=0.88$. Subsequent tidal stripping could reduce the mass of the PBH--halo system and potentially bring it into agreement with the inferred mass model. Thus, the PBH scenario remains an interesting possibility, although a more quantitative assessment is needed to determine its viability.
}

\section{Conclusion}
\label{sec:con}

We have investigated SIDM and CDM interpretations of the million-solar-mass lensing perturber JVAS B1938+666-$\mathcal{V}$ in light of its latest inferred mass model, which contains an unresolved point-like component embedded within an extended mass envelope. We showed that such an inner structure is a characteristic prediction of the SIDM scenario, in which an ultra-dense core emerges within a smooth halo profile as the system evolves into the deep gravothermal collapse phase. In the CDM interpretation, the perturber is a tidally stripped remnant of a dark matter halo hosting a central black hole. In this scenario, the black hole plus a central density spike {\blue induced} by its deep gravitational potential give rise to the dense inner region. After substantial tidal mass loss from the progenitor halo, the resulting remnant can reproduce the inferred mass profile of the perturber. However, this scenario requires both early formation of the massive progenitor and extreme infall and orbital evolution histories. It will be important to investigate whether such initial conditions and orbits can arise in realistic cosmological environments. Moreover, future measurements with even higher spatial resolution could provide a more stringent test of the different interpretations. More broadly, probing the nature of dark matter through strong-lensing observations will become increasingly important with data from ongoing surveys~\cite{Nierenberg:2023tvi,Keeley:2025oig,Gilman:2025fhy,Nierenberg:2026tma} and upcoming facilities~\cite{Wedig:2025idn,LSSTDarkMatterGroup:2019mwo}. It is therefore important to develop robust theoretical predictions for the abundance, internal structure, and evolution of low-mass dark matter perturbers in different dark matter scenarios.

\appendix

\section{Tidal mass loss of the CDM progenitor}
\label{sec:tidal}

We use the semi-analytic formalism developed in \cite{Du:2024sbt} to estimate the tidal mass loss,
\begin{equation}
	\label{eq-dMdt}
	\frac{d \Msub}{d t}=-\alpha_{\mathrm{s}} \frac{\Msub-\Msub\left(<r_{\mathrm{t}}\right)}{T_{\mathrm{loss}}},
\end{equation}
where $\Msub$ is the total mass and $\Msub(<r_t)$ is the mass enclosed in its tidal radius, both computed from the density in Eq.~\ref{eq-rhoall}. The tidal radius is obtained by solving 
\begin{equation}
	r_{\mathrm{t}}=\left(\frac{G \Msub\left(<r_{\mathrm{t}}\right)}{-\left.\frac{\mathrm{d}^2 \Phi}{\mathrm{~d} r^2}\right|_{\rsub}}\right)^{1 / 3},
\end{equation}
where the term in the denominator is the force gradient evaluated at the orbital radius $\rsub$. The characteristic timescale $T_{\rm loss}$ is approximated by the dynamical time within the tidal radius, $T_{\rm dyn} = 2\pi \sqrt{r_t^3 / 16 G \Msub(<r_t)}$, and $\alpha_{\mathrm{s}} = 3.93$ is a fitting parameter calibrated from simulations.

\begin{figure}
	\centering
	\includegraphics[scale=0.4]{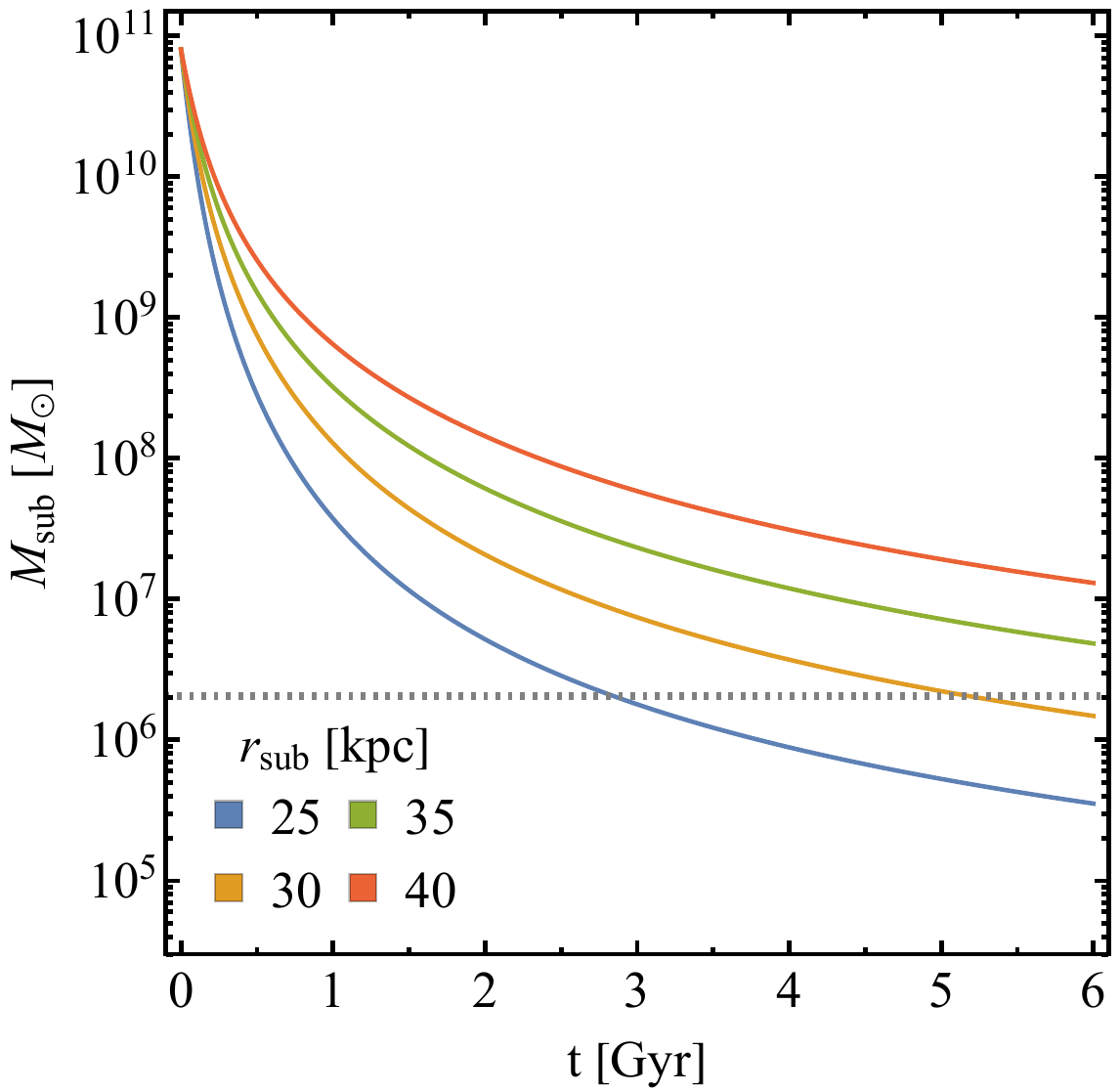}
	\caption{Time evolution of the subhalo mass assuming circular orbits $\rsub=25$, $30$, $35$, and $40\,{\rm kpc}$ from the center of the host galaxy. The horizontal dotted gray line denotes $\Msub = 2.1 \times 10^6 \, \Msun$ which best matches the mass model of the perturber.}
	\label{fig-orbit}
\end{figure}

To obtain the main lens potential $\Phi(r)$, we model the density distribution of the main lens galaxy as in~\cite{Powell:2025rmj}. The surface density is described by a power law
\begin{equation}
	\kappa(\xi) = \frac{3-\eta}{2}\left(\frac{R_{\rm E}}{\xi}\right)^{\eta-1}\Sigma_{\rm crit},
\end{equation}
where $\xi \equiv \sqrt{x^2 q+y^2/q}$ is the angular elliptical radius in arcsec \citep{Tajalli:2025qjx}, $q$ is the axis ratio, $\eta$ is the logarithmic slope of the three-dimensional density profile, $\Sigma_{\rm crit} = 1.5\times10^{11}\,\Msun\,{\rm arcsec^{-2}}$ is the critical surface density of the host galaxy. $R_{\rm E}$ is the Einstein radius 
\begin{equation}
	R_{\rm E} \equiv \left(\frac{\kappa_0\left(2-\frac{\eta}{2}\right) q^{(\eta-2) / 2}}{3-\eta}\right)^{1 /(\eta-1)},
\end{equation}
where $\kappa_0$ is a normalization parameter for surface density. We adopt $\kappa_0 = 0.54$ and $\eta = 1.86$ \citep{Tajalli:2025qjx}, and assume a spherical profile ($q=1$) for simplicity. The resulting three-dimensional density profile is 
\begin{equation}
	\rho_{\rm host} (r) = 1.19 \times10^9 \left(\frac{r}{{\rm kpc}}\right)^{-1.86}\Msun\,{\rm kpc^{-3}}.
\end{equation}
Figure~\ref{fig-orbit} shows the evolution of the subhalo mass obtained by solving Eq.~\ref{eq-dMdt}, assuming circular orbits with radii of $\rsub = 25$, $30$, $35$, and $40\,{\rm kpc}$. The horizontal gray dotted line marks $2.1\times10^6 \, \Msun$, corresponding to the mass of the matched dark matter profile. We find that the remnant mass falls below this threshold within $6\,{\rm Gyr}$ only for orbits with $\rsub \lesssim 30\,{\rm kpc}$.

\section*{acknowledgments}
We thank Demao Kong and Morgan Ohana for helpful discussions. This research was supported by the John Templeton Foundation under Grant ID \# 63599 and the U.S. Department of Energy under Grant No. DE-SC0008541. The opinions expressed in this publication are those of the authors and do not necessarily reflect the views of the funding agencies.

\section*{DATA AVAILABILITY}

The code used to perform the fluid simulations is publicly available at~\cite{data:2026}.
\bibliography{refbib-lensing}

\end{document}